\documentstyle[12pt]{article}
\title{\bf CORRELATION BETWEEN QUANTUM MECHANICS AND
CLASSICAL THEORY OF ROTATING ELECTRON MODELS AND POSSIBLE EXPERIMENT}
\author{V.N.Melekhin}
\date{17 January 1997}
\begin{document}
\maketitle

\begin{abstract}

It is shown that the point charge and magnetic moment of electron
produce together such a field that
total electromagnetic momentum has a component perpendicular
to electron velocity. As a result classical electron models, having
magnetic moment, move not along a straight line, if there is no
external force, but along a spiral, the space period and radius of
which are comparable with de-Broglie wave length.  Some other
surprising coincidences with quantum theory arises as a result of
calculation. An experiment is proposed for direct observation of
quantum or of new type electron delocalization.

\end{abstract}

\section{Introduction}

It is well known that electron spin and magnetic moment are pure
quantum electron characteristics having no classical analogs.  The 
main
reason for such a conclusion is that the classical equation of 
electron
motion can be derived from Shroedinger equation only in the limit of
$\hbar \to 0$ when spin disappeares from the theory. On the other 
hand,
if to introduce spin into classical theory, only new problems arises,
such as superlight charge velocity, and there is no correspondence 
with
quantum theory, as it was demonstrated as early as 1926 by Frenkel 
[1].

However, there is some problem. If to calculate the electromagnetic
momentum of a field generated by point electron charge together with
its point magnetic moment, then one can see that in the general
case this momentum has a component which is perpendicular to electron
velocity. Such a strange result sends us to study situation more
carefully using classical electron models. It has been found that this
component of momentum changes the classical motion radically. 
Moreover,
many coincidences with the quantum theory arises, to our astonishment.
Of course, all these coincidences can be occasional, so some
experiment is proposed to check a new approach.

\section{Motion of rotating electron models}

The equations for the motion of a classical electron model, having
intrinsic angular momentum and moving in external electric field
${\bf {\cal E}_{\circ}}$, are

$$
\frac{d{\bf P}}{dt} = e{\bf {\cal E}}_{\circ}, \; \;\;
\frac{d{\bf M}}{dt} = [{\bf P}{\bf v}], \eqno(1)
$$

where ${\bf P}$ is momentum, ${\bf M} = M_{\circ} \hat{\bf m}$
is angular momentum calculated
relative to the center of symmetry of a charge $e$, $\hat{\bf m}$
is the unit
vector issued from the center of symmetry and directed along the axis
of symmetry. The value of ${\bf P}$ is determined by the expression

$$
{\bf P} = \frac{U_{\circ}}{c^2} \left [ (1+\kappa) {\bf v} -
(3 \kappa -1) \hat{\bf m} (\hat{\bf m} {\bf v}) \right ],\eqno(2)
$$

where $U_{\circ}$ is the electromagnetic energy of a model,
the center of which
is at rest (for simplicity we count the charge to be of finite size 
and
the nonelectromagnetic mass to be equal to zero), ${\bf v}$
is a velocity of
the center of the charge ($v \ll c$), and $\kappa U_{\circ}$  is that
part of the total energy $U_{\circ}$ which is created by the 
components
of electric and magnetic fields being parallel to the axis of
symmetry.

The most important feature of eq.(2) is that there is a component of
${\bf P}$ being parallel to the vector $\hat{\bf m}$ and, hence,
in the general
case not parallel to the vector ${\bf v}$. As a result, the right side
of the second equation (1) is not equal to zero, and this changes
radically the character of the classical motion. Let us emphasize that
this perpendicular to ${\bf v}$ component $P_{\perp}$ arises not
due to the finite size of
discussed classical model but it is caused by the presence of "spin",
because, as it was mentioned earlier, for a point electron model,
having magnetic moment, the expression for the electromagnetic field
momentum is similar to eq.(2).  It is worth noting that the existence
of such a component $P_{\perp}$  was revealed even in paper [1],
however, the conclusion was made that this component vanishes because
the axis of symmetry must rotate, under the influence of some
"intrinsic" turning moment, in such a way that
$(\hat{\bf m}{\bf v}) \to 0$. Let us show this assumption to be 
untrue.

Using eqs.(1) and (2) one can derive the following equations for an
electron model motion

$$
\hspace{-5em}\frac{d{\bf v}}{dt}  =  \frac{e{\bf
{\cal E}_{\circ}}}{\mbox{m}_{\circ} (1+\kappa)} + \frac{(3\kappa-1)
e\hat{\bf m} (\hat{\bf m} {\bf {\cal E}_{\circ}})} {2
\mbox{m}_{\circ} (1-\kappa^2)}-
\frac{\mbox{m}_{\circ} (3 \kappa-1)^2}{M_{\circ} (1+\kappa)}
[\hat{\bf m} {\bf v}] (\hat{\bf m} {\bf v})^2 ,
$$

$$
\hspace{-5em}\frac{d\hat{\bf m}}{dt}  =
- \frac{\mbox{m}_{\circ}
(3 \kappa-1)}{M_{\circ} }
[\hat{\bf m} {\bf v}] (\hat{\bf m} {\bf v}), \eqno(3)
$$

where $\mbox{m}_{\circ}=\frac{U_{\circ} }{c^2}$.

It is significant that in some particular cases we used not only the
described method, when obtaining the equations (3), but some other 
one,
that uses more involved mathematics but it is more direct method. In
our paper [2] the formulae for a near field of a point charge, moving
with acceleration and having an arbitrary velocity, were first 
obtained
in an explicit form as a dependence of the field on
coordinates. Using these expressions we determined the self-force and
the turning moment produced by self-forces. Setting these values equal
and opposite to external ones, we obtained the same equations (3). It
proves that eqs.(3) hold and they really provide the conservation of
momentum and angular momentum.

It follows from eqs.(3) that for free motion, when an external
field ${{\bf {\cal E}}_{\circ}=0}$, the center of a model moves
not along a straight line but
along a spiral that we shall name "free spiral". The vector
$\hat{\bf m}$ is also rotating about the axis of free spiral.
The radius $R_{s}$
of this spiral, the angular frequency $\Omega_{s}$ of rotation of a
model about the axis of a spiral and its space period $\lambda_{s}$ 
are
as follows:

$$
R_s  =  \frac{M_{\circ} \sqrt{1-{\hat m}_z^2}}
{\mbox{m}_e v_z}, \;\;\;
\Omega_s  =  \frac{\mbox{m}_e {\hat m}_z v_z^2}
{M_{\circ} ( G +{\hat m}_z^2)},\;\;\;
$$

$$
\lambda_s  =  v_z \frac{2 \pi}{\Omega_s},\;\;\;
G  =  \frac{2 (1-\kappa)}{3 \kappa -1}. \eqno(4)
$$

Here $\hat{m}_{z}$ and $v_{z}$ are the projections of a unit vector
$\hat{\bf m}$, that determines the direction of spin, and of velocity
${\bf v}$ on the axis of free spiral which we assume to be directed
along $z$-axis $(\hat{m}_{z},v_{z} \equiv \mbox{const})$, the value of
 $\mbox{m}_e$ is an effective mass of discussed model that is
determined by the following expressions:

$$
\mbox{m}_e = \mbox{m}_{\circ}
\frac{(1+\kappa) G}{G +{\hat m}_z^2}, \;\;\;
P_z = \mbox{m}_e v_z. \eqno(5)
$$

The $x,y$-oscillations of the center of a model relative to the axis 
of
free spiral let us call free oscillations.

Notice that such a curved motion is the only chance not to violate the
conservation laws in case when perpendicular component $P_{\perp}$
exists, which
is caused by "spin" angular momentum. It is very curious that for
$M_{\circ} \approx \hbar$
the radius $R_{s}$ of free spiral is $R_{s} \approx \lambda_{\circ}$
with $\lambda_{\circ}$ being de-Broglie wave length. It means that 
such
a classical electron model executes high-frequency transverse
oscillations and for this reason it is delocalized by the value
$\lambda_{\circ}$
independently on its own size, however small this size may be.  At the
same time the motion, averaged through a period of free oscillations,
is usual classical motion of a particle of mass $\mbox{m}_e$, as it
shows numerical calculation of eqs.(3) when the mean radius of
curvature is much greater than the free spiral radius $R_{s}$.

Consequently, the free oscillations of a model allows us to overcome
one of the most difficult problem of any theory using nonspherical
electron model, because for such a model the "transverse" mass is not
equal to the "longitudinal" one even at $v \to 0$. Let us remember
that for a spherical model these masses are equal to each other but
$M_{\circ} \approx \hbar$ only for superlight charge velocity [1],
whereas for nonspherical model there is no such problem.

\section{Other coincidences with quantum theory}

The mentioned delocalization of a classical model by the value
$\lambda_{\circ}$  is not
unique correspondence with the quantum theory. For further discussion 
let
us assume that

$$
M_{\circ} {\hat m}_z
 = \pm \frac{\hbar}{2}, \;\;\; {\hat m}_z =
\pm \sqrt{\frac{G}{3}},\eqno(6)
$$

i.e., the projection of the angular momentum ${\bf M}$ on the axis of
the free spiral we count to be equal to the electron spin projection.
Let us emphasize that this component of the angular momentum is the
only one because transverse components of "spin" are compensated by 
the
proper components of orbital momentum arising due to model rotation
about the axis of free spiral. It follows from eqs.(4) and (6) that
$\lambda_s=2\lambda_{\circ}$. As a result, resonant phenomena
take place, if our model interacts
with periodic electric field, when its space period coincides with
de-Broglie wave length, because after every half a period of free
oscillations one can change $\hat {\bf m}$ by $-\hat {\bf m}$,
and this does not change two first members of
the first equation (3) which determine the interaction of a model with
an external electric field.

If an external field changes monotonically but an electron model
executes some periodic motion, then again the resonance between the
oscillations along the averaged trajectory and the transverse
oscillations about this trajectory leads to a set of discrete energy
levels. In particular, if an external field is not too strong and for
this reason the free spiral is changed only slightly then the distance
between these levels is just the same as it determines by the
quasiclassical calculation because, as it follows from eqs.(4) and 
(6),
phase shift of free oscillations along the trajectory is strictly the
same as the quasiclassical $\psi$-function phase shift.

Moreover, if to calculate the electron model motion when electrical
field ${\cal E}_z$, linearly changed in $z$-direction, is as strong as
one likes, then energy spectrum has the same constant energy
separation as it takes place in quantum theory for harmonic 
oscillator.
Let us point out some additional correspondence of new theory with 
quantum
mechanics.

If to introduce formally the values
${\bigtriangleup p_{x,y}=\mbox{m}_e v_{x,y}}$
and to take into account eqs.(6) and the relations
$\bigtriangleup x=\bigtriangleup y=R_s$,
then $\bigtriangleup p_x \cdot \bigtriangleup x$ and
$\bigtriangleup p_y \cdot \bigtriangleup y$
are comparable with $\hbar$. It means that
delocalization of electron models, dictated by free oscillations,
formally obeys usual uncertainty principle.

The commonly accepted standpoint is that spin is a pure quantum
property which has no classical analog. If so, all described
coincidences between new theory and quantum one must be treated as
occasional. As it was mentioned, the main objection against using the
concept of spin in classical theory is the fact that the equations of
classical motion follow from the quantum theory only at $\hbar \to 0$,
when spin disappears from the theory. However, we can see that new
approach allows to determine classical trajectories not in this
limiting case but, due to averaging through free oscillations period, 
at
finite value of $\hbar$, if the radius of curvature   is much greater
than the radius of free spiral $R_s$. So, the main objection against
"classical" spin is eliminated, to our opinion.

One could see that the parameters of free spiral obey the "uncertainty
principle", hence, the radius of spiral $R_s$ is so small that usually
this spiral may not be observable. For this reason the existence of 
the
free spiral contradicts nor to the classical theory nor to the
experimental results. The quantum theory does not include the concept 
of
trajectories, so it is possible that here there is no contradiction 
too.
If to take into account that the possible existence of the free spiral
follows just from the conservation laws, may be, this result is
something more than a fixed idea?

\section{Discussion and "experimentum crucis"}

A set of new problems arises if to admit the existence of the
"classical" spin but, on the other hand, we could solve some difficult
old problems.  In quantum theory delocalization of the electron by the
value $\lambda_{\circ}$, where $\lambda_{\circ} \to \infty$ at
$v \to 0$ , ceases to be incomprehensible and obtains simple physical
meaning.
In classical theory the finite electron size, which exceeds
"classical" electron radius ($10^{-13}$cm), eliminates the well-known
problem of so called "run-away" solutions of Lorentz-Dirac equation
(see [3]).

Some other problems also could be solved in very unusual manner. For
example, high-energy electron-electron scattering, at first sight,
excludes any other model than point electron. However, this is  true
only for the spherical model. Let us calculate the model with
pure electromagnetic mass, which consists of thin charged ring, 
rotating
with the velocity of light. In this case greater radius $b$ is about
$10^{-11}$cm, whereas little radius $a$ is only $10^{-120}$cm.
Hence, two such models, placed in two parallel planes, could slip
at such little distance between the centers of these rings that it
would be indistinguishable from point electrons.

At first sight, not only point model but even extended model must
include nonelectromagnetic mass, because some attractive
nonelectromagnetic force must compensate coulomb repulsion. However,
let us examine this problem more carefully. When the distance between
two separated charges increases the mutual electric energy, decreasing
at such process, could convert into magnetic energy of each moving
charge, and the work of coulomb force measures this conversion. In 
case
of two charged parts of the same electron model the situation is quite
different. For example, in case of uniformly charged sphere such a
conversion is prohibited because the magnetic field is equal to zero 
at
any point from symmetry considerations when radial expansion takes
place. It happens so because the magnetic fields, produced by 
different
parts of a charge, are mutually cancelled.

Such interference of magnetic fields changes completely the concept of
a force that acts "inside" the electron. Mentioned annihilation of
magnetic fields at each space point could observe only for the 
spherical
model, but for any other model this interference also does not permit
the exact conversion of electric energy into magnetic one.
Consequently, "inside" the electron one can not introduce the force of
mutual coulomb repulsion as a measure of such energy exchange. It is
worth remembering Lorentz' idea that we probably make a mistake when
we try to apply our usual concept of a force to different parts of the
same electron (see paragraph 182 in monograph [4]). If so, why we can
not discuss the electron model with pure electromagnetic mass that 
does
not expand and moves in such a way that all conservation laws hold?

Of course, new approach gives us a set of new problems but their
discussion in this paper seems to be premature. We might discuss them
in more detail if any experimental results were obtained that could
support new concept. Probably, such an experiment is as follows. It
consists of observation of very slow electrons passage (with the 
energy
of $10^{-2}$eV) through a nuclear filter having holes of the
diameter $D \sim 10^{-6}$cm. For electrons, having so low energy
that $\lambda_{\circ}>D$, the passage of classical point models
would not change, the part
of real passed electrons must decrease gradually because of
diffraction, and new electron models, moving along the free spiral,
must demonstrate a drastic decrease in the number of passed electrons
when $R_s>\frac{D}{2}$.

\section{Conclusion}

A reader of this paper must not conclude that its author tried to
develop pure classical electron theory that could explain quantum
electron behavior. It is well known that there exist too many problems
which prevent us from such an attempt. Nevertheless, the commonly
accepted point of view that spin is pure quantum property, having no
classical analogy, seems to be too vigorous, and in this paper we 
tried
to use some new ideas to introduce spin into classical theory. In any
case, the above-mentioned challenge concerning with the 
electromagnetic
momenrum component, which is perpendicular to electron velocity, must
be investigated not only for classical model but for real electron 
too.

\end{document}